\documentclass[reprint,onecolumn,prd,superscriptaddress,floatfix,nofootinbib,showpacs,showkeys]{revtex4} 
\usepackage[T1]{fontenc}
\usepackage[utf8]{inputenc}
\usepackage{graphicx,amsmath,amssymb}
\usepackage[colorlinks, linkcolor=blue, citecolor=blue, urlcolor=blue]{hyperref}
\usepackage{dcolumn}
\usepackage{multirow}

\begin{document}

\title{ 
Accurate theoretical prediction on positron lifetime of bulk materials
}

\author{Wenshuai Zhang} 
\email{wszhang@mail.ustc.edu.cn} 
\affiliation{Department of Modern Physics, University of Science and Technology of China, Hefei 230026, China} 
\affiliation{State Key Laboratory of Particle Detection and Electronics, USTC, Hefei 230026, China} 

\author{Bingchuan Gu} 
\email{glacierg@mail.ustc.edu.cn} 
\affiliation{Department of Modern Physics, University of Science and Technology of China, Hefei 230026, China}
\affiliation{State Key Laboratory of Particle Detection and Electronics, USTC, Hefei 230026, China}

\author{Jiandang Liu}
\email{liujd@mail.ustc.edu.cn} 
\affiliation{Department of Modern Physics, University of Science and Technology of China, Hefei 230026, China}
\affiliation{State Key Laboratory of Particle Detection and Electronics, USTC, Hefei 230026, China}

\author{Bangjiao Ye} 
\email{bjye@ustc.edu.cn} 
\affiliation{Department of Modern Physics, University of Science and Technology of China, Hefei 230026, China}
\affiliation{State Key Laboratory of Particle Detection and Electronics, USTC, Hefei 230026, China}


\begin{abstract}
Based on the first-principles calculations,
we perform an initiatory statistical assessment on the reliability level of 
theoretical positron lifetime of bulk material. 
We found the original generalized gradient approximation (GGA) form of the enhancement 
factor and correlation potentials overestimates the effect of the gradient factor.
Furthermore, an excellent agreement between model and data with the difference 
being the noise level of the data is found in this work. 
In addition, we suggest a new GGA form of the correlation scheme which gives the best performance. 
This work demonstrates that a brand-new reliability level is achieved for the theoretical 
prediction on positron lifetime of bulk material and 
the accuracy of the best theoretical scheme can be independent on the type of materials. 
\end{abstract}


\maketitle

\section{Introduction}
During recent years positron annihilation spectroscopy (PAS) has become
a valuable method to study the microscopic structure of solids \cite{Tuomi13V85,Puska94V66} 
and gives detailed information on the electron density and momentum distribution
\cite{Makko14V89,Kontr12V85,Tang05V94} in the regions scanned by positrons. 
For a thorough understanding and interpretation of experimental results,
an accurate theory is needed. 
Exact many-body theory calculations on annihilation rate and scattering dynamics 
can be implemented 
for the positron in small atom or molecule system 
\cite{Mitro02V65,Griba10V82,Green14V90}, 
but is time-consuming for the positron in large many-electron system. 
Based on the density functional theory (DFT) \cite{Kohn65V140}, 
a full two-component self-consistent scheme \cite{Niemi85V32,Puska95V52} 
has been developed for calculating positron states in solids. 
Especially in bulk material where the positron is delocalized 
and does not affect the electron states, 
the full two-component scheme can be reduced without losing accuracy 
to the conventional scheme \cite{Niemi85V32,Puska95V52} 
in which the electronic-structure is determined by usual one-component formalism. 
However, there are various kinds of approximations on electron-positron correlation
can be adjusted within this calculations. 
To improve the analyses of experimental data \cite{Wikto14V89,Campi12V14}, 
we should find out which approximations are more credible 
to predict the positron lifetimes. 
Thus, in this short paper, we focus on probing the reliability level of 
these approximations for calculating the positron lifetimes in bulk materials.

Recently, Drummond et al. \cite{Drumm11V107} made the most accurate calculations 
for a positron in a homogeneous electron gas by using Quantum Monte Carlo (QMC) 
method and gave a smaller enhancement factor 
compared with the popular expression \cite{Boron86V34}. 
Very recently, Kuriplach and Barbiellini \cite{Kurip14V89,Kurip14V505} 
implemented multiple calculations of positron-annihilation characteristics in solid 
based on the local density approximation (LDA) or generalized gradient approximation (GGA) 
forms of the enhancement factor and correlation potential provided by the 
perturbed hypernetted chain (PHC) calculation \cite{Stach93V48,Boron10V55}
or reparameterized from Drummond et al.'s QMC results.
Their results showed that the recent two GGA forms of the 
correlation schemes are needed to improve the calculated positron lifetimes. 
But it's hard to clearly judge and distinguish the reliability level of 
these two GGA models based on one by one comparisons with a small number of materials. 
For more recent studies on the calculations of positron lifetimes, 
see Refs. \cite{Boron10V55,Taken08V77}.

In this paper, we investigate nine LDA/GGA correlation schemes 
containing a new GGA form for positron lifetime calculations 
based on the full-potential linearized augmented plane-wave (FLAPW) plus 
local orbitals approach \cite{Sjoes00V114} for accurate 
electronic-structure calculations. 
The experimental data used in this work are composed of many observed values 
of materials more than twice as much as previous works
\cite{Kurip14V89,Kurip14V505,Boron10V55,Taken08V77}. 
To take into account the fact that the materials having more credible 
experimental values should play more important roles in these assessments,
the measurement errors of these experimental values are assumed being Gaussian 
and then estimated by the standard deviations of collected observed values 
from different literatures and/or groups as in Ref. \cite{Ito08V104}. 
Furthermore, five subsets are structured depending on the number of 
observed values of each material to make a subtler probe. 
By utilizing this data, we do the initiatory numerical and statistic assessment
on the reliability level of various LDA and/or GGA correlation schemes 
for positron lifetime calculations.

This paper is organized as follows: In Sec. II, we give a
brief description of the models considered here as
well as the analysis methods we used. In Sec. III, we introduce 
the experimental data on positron lifetime used in this work. 
In Sec. IV, we give the results and make some discussion based on 
the visualized and statistic analyses. 
In Sec. V, we make some conclusions of this work. 
In addition, we present a appendix with a table 
listing all calculated theoretical lifetimes.

\section{Theory and methodology}

\subsection{Theory} 
The positron lifetime which is the inverse of positron annihilation rate
can be obtained by the following equations \cite{Boron86V34} in bulk materials,
\begin{equation} 
\label{eq:tau} 
\tau_{e+ }=\frac{1}{\lambda},\ \ \lambda=\pi r_{0}^{2}c\int d\vec{r}n_{e- }(\vec{r})n_{e+ }(\vec{r})\gamma(n_{e- }),
\end{equation} 
where $r_{0}$ is the classical electron radius, $c$ is the speed
of light, and $\gamma(n_{e- })$ is the enhancement factor 
arising from the contact pair-correlation between positron and electrons. 
For a perfect lattice, the conventional scheme is still accurate as 
in this case the positron density is delocalized and 
vanishingly small at every point thus does not 
affect the bulk electronic-structure \cite{Puska95V52,Boron86V34}. 
So in this paper the electronic density $n_{e- }(\vec{r})$ 
were calculated without considering the perturbation by positron
based on the FLAPW approach \cite{Sjoes00V114} 
which is regarded as the most accurate method for electronic-structure calculations. 
The total potential sensed by positron is composed of the Coulomb potential 
and the correlation potential \cite{Boron86V34} between electrons and positron. 
Then, the positron density can be determined by 
solving the Kohn-Sham equation \cite{Kurip14V89}.  
\begin{table}[htb] 
\footnotesize
\caption{
\label{tab:enhfct}
Nine parameterized LDA/GGA correlation schemes. 
}
\begin{tabular}{ l r r r r r r r }
\hline
\hline
$\gamma$ & $a_2$ & $a_3$ & $a_{3/2}$ & $a_{5/2}$ & $a_{7/3}$ & $a_{8/3}$ & $\alpha$  \tabularnewline 
\hline
\\[-2ex]
BNLDA  & -1.26 	& $1/6-1/6\epsilon_{\infty}$ & 0.8295 & 0.3286 & 0 & 0 & 0  \tabularnewline
APLDA  & -0.0742 &  1/6    & 0 & 0 & 0 & 0 & 0  \tabularnewline 
APGGA  & -0.0742 &  1/6    & 0 & 0 & 0 & 0 & 0.22  \tabularnewline 
PHCLDA & -0.137 &  1/6    & 0 & 0 & 0 & 0 & 0   \tabularnewline 
PHCGGA & -0.137 &  1/6    & 0 & 0 & 0 & 0 & 0.10  \tabularnewline 
QMCLDA & 8.6957 &  0.1737 & -3.382 & 0 & -7.37 & 1.756 & 0 \tabularnewline 
QMCGGA & 8.6957 &  0.1737 & -3.382 & 0 & -7.37 & 1.756 & 0.063 \tabularnewline 
fQMCLDA  & -0.22 &  1/6    & 0 & 0 & 0 & 0 & 0  \tabularnewline 
fQMCGGA  & -0.22 &  1/6    & 0 & 0 & 0 & 0 & 0.05   \tabularnewline 
\hline 
\hline 
\end{tabular} 
\end{table}
The forms of enhancement factor and correlation potential can be divided 
into two categories: the LDA and the GGA. 
Within the LDA, the corresponding correlation potential $V_{corr}^{\text{LDA}}$ 
given by Ref. \cite{Boron86V34} is used. 
Within the GGA, the corresponding correlation potential takes the form \cite{Barbi95V51,Barbi96V53} 
$V_{corr}^{ \text{GGA} }=V_{corr}^{\text{LDA} }e^{-\alpha\epsilon/3}$, 
here $\alpha$ is an experiential parameter, and $\epsilon$ is defined as 
$\epsilon=|\nabla\ln(n_{e- })|^{2} / q_{\text{TF}}^{2}$,
($q_{\text{TF}}^{-1}$ is the local Thomas-Fermi screening length). 
We investigated eight existing forms of the enhancement factor and correlation potential
marked by BNLDA\cite{Puska89V39}, APLDA\cite{Barbi95V51}, APGGA\cite{Barbi95V51,Barbi96V53}, 
PHCLDA\cite{Stach93V48},PHCGGA\cite{Boron10V55}, QMCLDA\cite{Drumm11V107},
fQMCLDA\cite{Kurip14V89} and fQMCGGA\cite{Kurip14V89}, plus a new GGA form QMCGGA.
All forms of the enhancement factor can be parameterized by the following equation, 
\begin{eqnarray} 
\gamma =
& 1 + (1.23r_{s} + a_{2}r_{s}^{2} + a_3 r_{s}^{3}  + a_{3/2} r_{s}^{3/2} 
& +\ a_{5/2}r_{s}^{5/2} + a_{7/3}r_{s}^{7/3} + a_{8/3}r_{s}^{8/3} ) 
e^{-\alpha\epsilon} , 
\end{eqnarray} 
here $r_{s}$ is defined by $r_{s}=(3/4\pi n_{e-})^{1/3}$, 
and the values of the parameters 
$a_{2}, a_3, a_{3/2}, a_{5/2}, a_{7/3}, a_{8/3}, \text{and } \alpha$ 
are listed in Table \ref{tab:enhfct} according to specific kind of the correlation scheme.
The QMCGGA form proposed in this work is derived from the original 
QMCLDA parametrization introduced by Drummond et al. \cite{Drumm11V107}, 
instead of the APLDA parametrization used to fit the QMCLDA results 
within the fQMCLDA and the fQMCGGA. 
The adoption of the QMCLDA parametrization is due to the fact that 
the existence of positive $a_{8/3}$ term and the lager $a_3$ term 
lead to a much lager enhancement in the high $r_s$ rigion compared with the fQMCLDA form. 
Nevertheless, the difference between QMCLDA and fQMCLDA at low $r_s$ ($r_s<6$) is minor 
and the fitted parameter $\alpha$ is only slightly changed from 0.05 to 0.063, 
which will result in similar lifetime values for most materials.

\subsection{Computational details}
In practice of this work,
The WIEN2k code \cite{Blaha01Voo} was used for the FLAPW electronic-structure calculations.
The PBE-GGA approach \cite{Perde96V77} was adopted for electron-electron exchange-correlations, 
the total number of k-points in the whole Brillouin zone (BZ) was set to 3375, 
the default values of muffin-tin radius were used,
and the self-consistency was achieved up to 
both levels of 0.0001 Ry for total energy and 0.001 e for charge distance. 
To obtain the positron-state, the three-dimensional Kohn-Sham equation
was solved by the finite-difference method while the unit cell of each material 
was divided into about 10 mesh spaces per $bohr$ in each dimension. 
All important variable parameters were checked carefully to achieve that the 
computational precision of lifetime values are at most the order of 0.1 ps.

\begin{table}[!htb]
\caption{ \label{exp-data} 
The number of observed values $n_{\textbf{obs} }$, 
the high-frequency dielectric constant $\epsilon_{\infty}$, 
the experimental values of lifetime $\tau_{\textbf{exp} }$ alongwith 
the corresponding mean value ${\tau}_{\textbf{exp} }^{*}$, 
and the standard deviation $\sigma_{\textbf{exp} }$ for each material in this work. 
The high-frequency dielectric constants not listed here 
will be set to $\infty$ as in Ref. \cite{Campi07V19}. 
}
\resizebox{0.66\textwidth}{!}{
\begin{tabular}[b]{ l l r r r r }
\hline 
\hline 
\rule[.2ex]{0pt}{2ex} 
$n_{\textbf{obs} }$ & Material & $\epsilon_{\infty}$ & $\tau_{\textbf{exp} }$ & ${\tau}_{\textbf{exp} }^{*}$ & $\sigma_{\textbf{exp} }$\tabularnewline
\cline{1-6}  
\rule[.25ex]{0pt}{2.5ex} 
%
1 & Li & - & 291 \cite{Weisb67V154}  & 291 & -\tabularnewline
 & Na & - & 338\cite{Weisb67V154} & 338 & -\tabularnewline
 & K & - & 397\cite{Weisb67V154} & 397 & -\tabularnewline
 & Rb & - & 406\cite{Weisb67V154} 	& 406 & -\tabularnewline
 & Cs & - & 418\cite{Weisb67V154} 	& 418 & -\tabularnewline
 & Ga & - & 198\cite{Campi03V213-215} 	& 198 & -\tabularnewline
 & Cr & - & 120 \cite{Campi03V213-215}	& 120 & -\tabularnewline
 & Sc & - & 230\cite{Welch76V77} 	& 230 & -\tabularnewline
 & Y & - & 249\cite{Campi03V213-215} 	& 249 & -\tabularnewline
 & TiO & - & 140 \cite{Valee07V75}	& 140 & -\tabularnewline
 & GaN & 5.4\cite{Puska89V39} & 180\cite{Puska89V39}  	& 180 & -\tabularnewline
 & ZnS & 5.4\cite{Plaza94V6} & 230\cite{Plaza94V6}  	& 230  & -\tabularnewline
 & ZnTe & 7.28\cite{Plaza94V6} & 266\cite{Plaza94V6} 	& 266 & -\tabularnewline
 & HgTe & 15\cite{Plaza94V6} & 274\cite{Geffr86Voo} & 274 & -\tabularnewline
 & ZnSe & 5.8\cite{Plaza94V6} & 240\cite{Plaza94V6}  	& 240  & -\tabularnewline
 & PbSe & 22.9\cite{Zemel65V140} & 220\cite{Polit93V131} & 220 & -\tabularnewline
 & NiO & 5.7\cite{Gieli65V36} & 110\cite{Barbi91V3} 	& 110 & -\tabularnewline
\\
%
2 & Be & - & 137 \cite{Campi03V213-215} 142\cite{Campi03V213-215}  & 139.5 & 3.535\tabularnewline
 & Pt & - & 99\cite{Campi03V213-215} 128\cite{Campi03V213-215} 	& 113.5  & 20.51\tabularnewline
 & Co\_alpha & - & 118\cite{Welch76V77} 119\cite{Campi03V213-215} 	& 118.5 & 0.707\tabularnewline
 & Tl & - & 226\cite{Campi03V213-215} 229\cite{Campi03V213-215} 	& 227.5  & 2.121\tabularnewline
 & TiC & - & 155\cite{Bisi85V5} 160\cite{Rempe93V5} 	& 157.5  & 3.535\tabularnewline
 & SiC\_3C & 6.52\cite{Puska89V39} & 138\cite{Braue96V54} 140\cite{Kawas98V67} 	& 139.0  & 12.02\tabularnewline
 & GaP & 9.1\cite{Puska89V39} & 223\cite{Dlube85V49} 225\cite{Danne82V15} 	& 224.0  & 1.414\tabularnewline
 & InAs & 12.3\cite{Puska89V39} & 247\cite{Danne82V15} 257\cite{Dlube86V7} 	& 252.0  & 7.071\tabularnewline
\\
%
3 & InSb & 15.7 \cite{Puska89V39} & 258\cite{Puska89V39} 280\cite{Gupta88V23} 282\cite{Danne82V15} 	& 273.3  & 13.32  \tabularnewline
 & GaSb & 14.4 \cite{Puska89V39} & 247\cite{Danne82V15} 260\cite{Puska89V39} 260\cite{Puska89V39} & 255.7  & 7.505 \tabularnewline
 & Bi & - & 227.5\cite{Campi03V213-215} 238\cite{Campi03V213-215} 241\cite{Campi03V213-215} 	& 235.5  &  7.088\tabularnewline
 & Zr & - & 165\cite{SeegeooVoo} 163\cite{Campi03V213-215}  165\cite{Campi03V213-215}  	& 164.3  & 1.154\tabularnewline
 & C\_diamond & 5.66\cite{Puska89V39} & 97.5\cite{Campi03V213-215}  107\cite{Campi03V213-215}  115\cite{Danne82V15}  & 106.5  & 8.760\tabularnewline
 & W & - & 105\cite{SeegeooVoo} 102\cite{Campi03V213-215} 116\cite{Campi03V213-215} 	& 107.7  & 7.371\tabularnewline
\\
%
4 & Pd & - & 96\cite{SeegeooVoo} 120\cite{Campi03V213-215} 108\cite{Campi03V213-215} 118\cite{Campi03V213-215} 	& 110.5  & 11.00\tabularnewline
 & V & - & 120\cite{Campi03V213-215} 130\cite{SeegeooVoo} 123\cite{Campi03V213-215} 130\cite{Campi03V213-215} 	& 125.5  & 5.058\tabularnewline
 & SiC\_6H & 6.52\cite{Puska89V39} & 141\cite{Braue96V54} 140\cite{Henry03V67} 146\cite{Valee07V75} 144\cite{Valee07V75} 	& 142.8 & 2.753\tabularnewline
 & MgO & 3.0\cite{Puska89V39} & 130\cite{Mizun04V445-446} 166\cite{Valee07V75} 152\cite{Valee07V75} 155\cite{Barbi91V3} 	& 150.8  & 15.09\tabularnewline
\\
%
5 & Si & 11.9\cite{Campi03V213-215} & 216.7\cite{Campi03V213-215} 218\cite{Campi03V213-215} 218\cite{Campi03V213-215} 222\cite{Campi03V213-215} 216\cite{Campi03V213-215} & 218.1 & 2.323\tabularnewline
 & Ge & 16.0\cite{Puska89V39} & 220.5\cite{Campi03V213-215} 230\cite{Campi03V213-215} 230\cite{Campi03V213-215} 228\cite{Campi03V213-215} 228\cite{Campi03V213-215} 	& 227.3  & 3.931\tabularnewline 
 & Mg & - & 225\cite{SeegeooVoo} 225\cite{Campi03V213-215} 220\cite{Campi03V213-215} 238\cite{Campi03V213-215} 235\cite{Campi03V213-215} & 228.6 & 7.569\tabularnewline
 & Al & - & 160.7\cite{Campi03V213-215} 166\cite{Campi03V213-215} 163\cite{Campi03V213-215} 165\cite{Campi03V213-215} 165\cite{Campi03V213-215} & 163.9  & 2.114\tabularnewline
 & Ti & - & 147\cite{SeegeooVoo} 154\cite{Campi03V213-215} 145\cite{Campi03V213-215} 152\cite{Campi03V213-215} 143\cite{Campi03V213-215} & 148.2 & 4.658\tabularnewline
 & Fe & - & 108\cite{Campi03V213-215} 106\cite{Campi03V213-215} 114\cite{Campi03V213-215} 110\cite{Campi03V213-215} 111\cite{Campi03V213-215} & 109.8 & 3.033\tabularnewline
 & Ni & - & 109.8\cite{Campi03V213-215} 107\cite{Campi03V213-215} 105\cite{Campi03V213-215} 109\cite{Campi03V213-215} 110\cite{Campi03V213-215} & 108.2 & 2.127\tabularnewline
 & Zn & - & 148\cite{SeegeooVoo} 153\cite{Campi03V213-215} 145\cite{Campi03V213-215} 154\cite{Campi03V213-215} 152\cite{Campi03V213-215} & 150.4  & 3.781\tabularnewline
 & Cu & - & 110.7\cite{Campi03V213-215} 122\cite{Campi03V213-215} 112\cite{Campi03V213-215} 110\cite{Campi03V213-215} 120\cite{Campi03V213-215} & 114.9  & 2.514\tabularnewline
 & Nb & - & 119\cite{Campi03V213-215} 120\cite{Campi03V213-215}  122\cite{Campi03V213-215} 122\cite{Campi03V213-215} 125\cite{Campi03V213-215} & 121.6  & 2.302\tabularnewline
 & Mo & - & 109.5\cite{Campi03V213-215} 103\cite{Campi03V213-215} 118\cite{Campi03V213-215} 114\cite{Campi03V213-215} 104\cite{Campi03V213-215} & 109.7  & 6.418\tabularnewline
 & Ta & - & 116\cite{SeegeooVoo} 122\cite{Campi03V213-215} 120\cite{Campi03V213-215} 125\cite{Campi03V213-215} 117\cite{Campi03V213-215} & 120.0  & 3.674\tabularnewline
 & Ag & - & 120\cite{Campi03V213-215} 130\cite{Campi03V213-215} 131\cite{Campi03V213-215} 133\cite{Welch76V77}  131\cite{SeegeooVoo}	& 129.0  & 5.147\tabularnewline
 & Au & - & 117\cite{Campi03V213-215} 113\cite{Campi03V213-215} 113\cite{Campi03V213-215} 117\cite{Campi03V213-215} 123\cite{Campi03V213-215} & 116.6  & 4.098\tabularnewline
 & Cd & - & 175\cite{SeegeooVoo} 184\cite{Campi03V213-215} 167\cite{Campi03V213-215} 172\cite{Campi03V213-215} 186\cite{Campi03V213-215} & 176.8  & 8.043\tabularnewline
 & In & - & 194.7\cite{Campi03V213-215} 200\cite{Campi03V213-215} 192\cite{Campi03V213-215} 193\cite{Campi03V213-215} 189\cite{Campi03V213-215} & 193.7  & 4.066\tabularnewline
 & Pb & - & 194\cite{SeegeooVoo} 200\cite{Campi03V213-215} 204\cite{Campi03V213-215} 200\cite{Campi03V213-215} 209\cite{Campi03V213-215} & 201.4  & 5.550\tabularnewline
 & GaAs & 10.9\cite{Puska89V39} & 231.6\cite{Wang00V177} 231\cite{Saari91V44} 230\cite{Polit97V55} 232\cite{Dlube87V42} 220\cite{Danne84V30} & 228.9  & 5.043\tabularnewline
 & InP & 9.60\cite{Puska89V39} &  241\cite{Belin98V58} 240\cite{Chen98V66} 247\cite{Puska89V39}  242\cite{Dlube86V7} 244\cite{Dlube85V46} & 242.8  & 2.775\tabularnewline
 & ZnO & 3.9\cite{Yoshi97V36} & 153\cite{Mizun04V45} 159\cite{Braue06V74} 158\cite{Uedon03V93} 161\cite{Brunn01V363-365} 171\cite{Tuomi03V91} & 160.4  & 6.618\tabularnewline
 & CdTe & 7.2\cite{Puska89V39} & 284\cite{Plaza94V6} 285\cite{Gely-93V80} 285\cite{Peng00V3} 289\cite{Geffr86Voo} 291\cite{Danne82V15} & 286.8  & 3.033\tabularnewline
\hline 
\hline 
\end{tabular} 
 }
\end{table}

\subsection{Model comparison}
To make a comparison between different models, an appropriate criterion
must be chosen. The popular one is the root mean squared deviation (RMSD)
which is defined as the square root of the mean of the squared deviation between 
experimental and theoretical results. 
Beyond this, a comprehensive statistical analysis should be employed where 
the credibility of observed lifetimes can be estimated by the standard deviations. 
Therefore, based on the chi-squared analysis, we also adopted 
$\chi^{2}/dof =  \sum^{N}_{i=1} [
(X_{i}^{\textbf{exp} }-X_{i}^{\textbf{theo} })^{2} / \sigma_i^2 N] $
as another selection criterion for different models and datasets, 
where $\sigma_{i}$ is the standard deviation 
of experimental value for each material, 
and the $dof$ (degree of freedom) is set to N 
(the size of corresponding dataset) 
since the parameters of each model are fixed in this work. 
In addition, the \textit{p}-value corresponding to each $\chi^{2}$ 
is much more meaningful to explore the agreement level of 
theoretical models and experimental data. 
From the above definitions, one can see that the experimental data 
favor models producing lower (higher) values 
of the RMSD and/or $\chi^{2}/dof$ (\textit{p}-values). 
Especially, the models with \textit{p}-value $< 0.01$ 
are most likely rejected by current collected data.

\section{Experimental data}
We gathered up to five recent observed values from different literatures and/or groups 
for 56 materials to compose a more reliable experimental dataset 
which consequently lead to more credible results
comparing with previous works \cite{Kurip14V89,Kurip14V505,Boron10V55,Taken08V77}.
All the data used in this work are listed in Table \ref{exp-data} 
and collected basically within the standard suggested in Ref. \cite{Campi07V19}. 
Older experimental data were avoided being adopted, 
and only one observed value is used for these materials 
whose experimental values are all measured before 1975. 
It's hard to choose a quantity to estimate their measurement errors. 
In our previous work \cite{Zhang14V2}, 
only the statistical error of single measurement is used in the model selection criterion. 
But the deviations of results between different groups 
are usually much larger than the statistical errors, 
even when just the recent and reliable measurements are considered. 
That is, the systematic error is the dominant factor 
and the sole statistical error is far from enough. 
However, the systematic error is difficult to derive from single experimental result. 
The time resolution of related measurement is a correlative quantity 
but is nearly at the same level in recent years and insufficient, 
since the quality of material sample is another significant factor which can not be neglected. 
Unfortunately, the definite effect of sample defect 
caused by unintentional doping is imponderable.  
In this paper, instead of focusing on the uncertain error of single experiment, 
we evaluate the reliability level of average experimental values of each material. 
As usual, we assume that the distribution of 
observed values from repeated measurements is Gaussian
and the systematic errors tend to be cancelled as expected in Ref. \cite{Campi07V19}, 
Hence, the measurement uncertainties of related materials 
can be estimated by the standard deviations of collected observed values 
from different literatures and/or groups as in Ref. \cite{Ito08V104}. 
It is reasonable since the materials with larger scattering observed lifetimes 
is more debatable and should play less important roles in these assessments. 
To make a subtle probe, five subsets were structured depending on the number of observed values:
(a) subset A includes the overall 56 materials,
(b) subset B includes 39 materials having at least two observed values,
(c) subset C includes 31 materials having at least three observed values,
(d) subset D includes 25 materials having at least four observed values,
(e) subset E includes 21 materials having five observed values. 
It should be noted that, 
for a more comprehensive assessment including positron lifetimes in defects,  
the calculations based on the full two-component scheme need to be performed. 
However, in this work, we focus on testing the calculation methods 
by using lifetime data of bulk materials because of the insufficient observed values 
together with their large scattering for vacancy lifetimes and the fact that 
the conventional scheme is strictly accurate for bulk-lifetime calculations.

\section{Results and discussion}

\subsection{Visualized analyses} 

\begin{figure*}[!htb]
\includegraphics[width=1.0\textwidth]{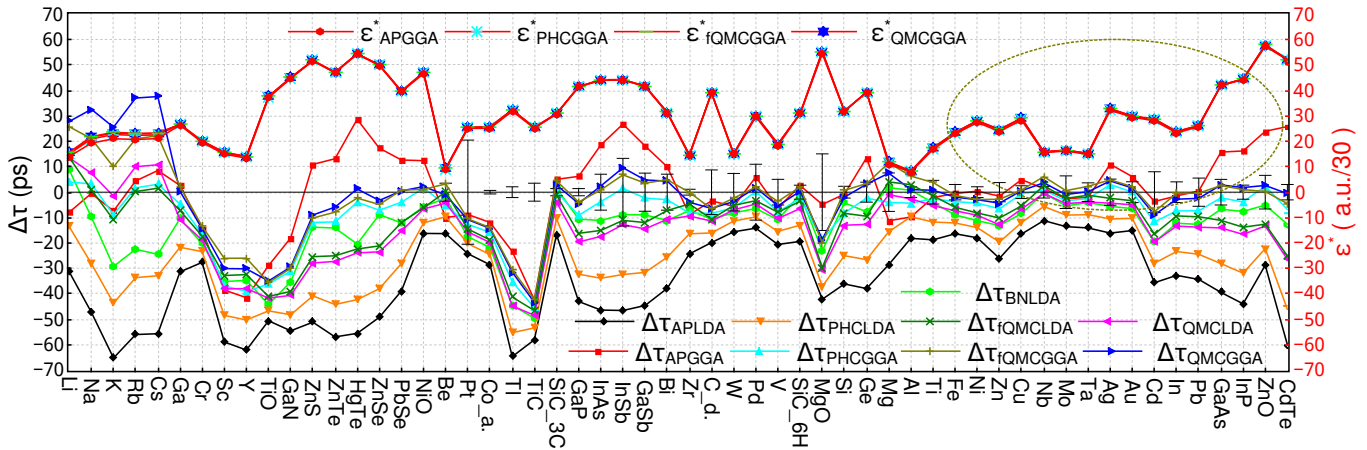} 
\caption{
\label{fig:ep-dt-FLAPW}
The deviations of the theoretical results based on various approximations 
from the experimental values $\Delta\tau=\tau_{\text{theo}} - \tau^{*}_{\text{exp}}$ 
alongwith the standard deviation of experimental values for each material.
To make a comparison, 
the mean electron-density gradient factor $\epsilon^{*}$ sensed by positron 
for four GGA forms based on the FLAPW calculations are also plotted. 
} 
\end{figure*}
We firstly give visualized comparisons between experimental values and 
theoretical results based on different correlation schemes.
All the detailed numerical theoretical results can be found in appendix.
The difference between theoretical results and experimental data 
along with the standard deviations of observed values for all materials 
are plotted in Fig. \ref{fig:ep-dt-FLAPW}.
From this figure, 
the scattering regions of calculated lifetimes 
by different forms of the enhancement factor 
are found much larger in the semiconductor systems with bonding states 
compared with those in pure metal systems excluding the alkali metals. 
This means the positron lifetimes of semiconductors and alkali metals 
have more sensitivity to different forms of the enhancement factor. 
For the GGA approaches, we also plot the average electron-density gradient factor
defined by
$\epsilon^{*} =\int \epsilon^{e-}_{ \text{FLAPW} } \cdot n^{e+}_{\text{GGA}} \cdot d\vec{r}$  
in Fig. \ref{fig:ep-dt-FLAPW}. 
This figure shows that the four GGA approaches give 
almost the same values of $\epsilon^{*}$.
Furthermore, the most interesting phenomenon in Fig. \ref{fig:ep-dt-FLAPW} is that 
a clear positive correlation between $\epsilon^{*}$ and $\Delta \tau_{\text{APGGA}}$ 
is found especially in the area of the more credible subset E
as marked with olive dashed ellipse.
Considering that the presence of electron-density gradient term $\epsilon$ decreases 
the enhancement factor $\gamma(n_{e- })$ in Eq. \eqref{eq:tau} and then increases the lifetime,
this phenomenon actually indicates that the original GGA form APGGA (with $\alpha=0.22$) 
overestimates the contribution of the gradient term  
while for other GGA forms no assured overestimation is appeared.

\subsection{Numerical and statistic analyses}

\begin{figure} [!htb]
\includegraphics[width=11cm]{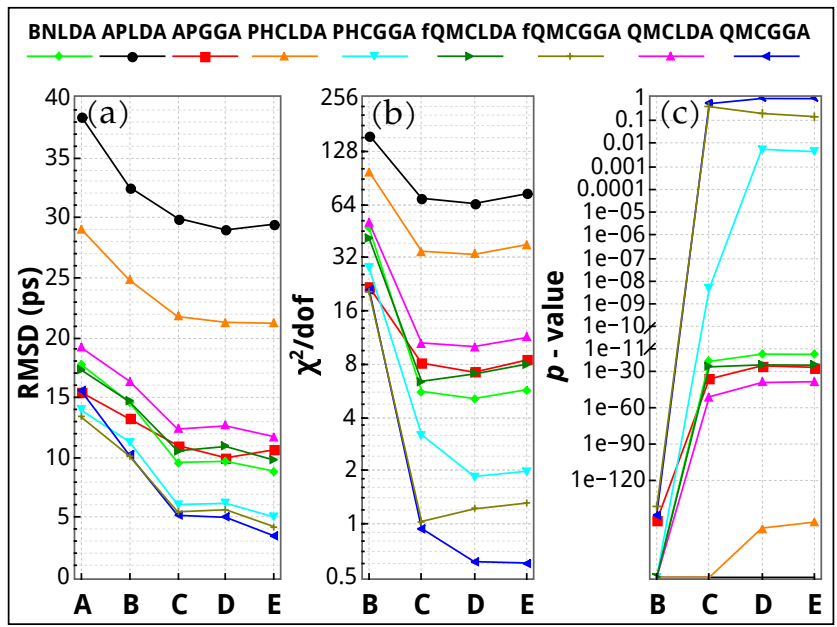} 
\caption{ 
\label{fig:stats} 
The RMSD (a), $\chi^2/dof$ (b) and \textit{p}-value (b) 
versus the subsets of data based on different forms 
of the correlation schemes. 
} 
\end{figure}

To make a precise and numerical assessment, we plot the values of RMSD 
for different approaches and subsets of data in Fig. \ref{fig:stats}(a). 
Before going on, 
we mention again that the subsets 
having more observed values for each material are more credible, that is, 
the subset E should play the most important role in the following discussions. 
It is reasonable to see that, 
with increasing the index of subset from A to E, 
the RMSDs for most forms of the enhancement factor reach to stable values 
except three recent GGA forms: PHCGGA, fQMCGGA and QMCGGA. 
This fact implies that these GGA forms may have the ability to 
give smaller values of RMSD with more experimental data in the future. 
Among those forms of enhancement factor based on recent QMC results \cite{Drumm11V107}, 
the fQMCLDA fitted by the Kuriplach and Barbiellini gives a smaller RMSD than 
the original one QMCLDA while both of them are not good enough. 
On the contrary, the QMCGGA form proposed in this work 
performs better than the fQMCGGA form and produces the best RMSD. 
Therefore, we will focus on discussing QMCGGA and QMCLDA forms 
instead of the reparameterized forms fQMCLDA and fQMCGGA. 
Several other phenomena can be found in this figure. 
Firstly, 
the QMCLDA form of the enhancement factor 
makes large progress compared with the older forms APLDA and PHCLDA, 
and the gradient correction to these LDA forms are still needed to 
give much better results which is consistent with previous study \cite{Kurip14V89}.
Secondly, 
the phenomenon that the best LDA form (BNLDA) performs 
a little better than the original GGA form (APGGA) 
within the exact self-consistent electronic calculation (FLAPW) 
confirms previous works \cite{Taken08V77}. 
There, they claimed that beyond LDA methods are not needed to reach a well 
level of agreement at that time when accurate band-structure methods (FLAPW) are used. 
However, our results further show that the recent three GGA approaches (PHCGGA, 
fQMCGGA and QMCGGA) make significant improvement on the agreement 
between theoretical and experimental lifetimes compared with the best LDA approach. 
Finally, 
based on the FLAPW method, the QMCGGA and PHCGGA forms give a best value of 3.46 ps
and a passable value of 5.04 ps for RMSD, respectively.
Because the more closer to zero the harder to reduce RMSD, 
the decrease of RMSD from 5.04 ps to 3.46 ps maybe noteworthy. 
To investigate this problem, a further statistic analysis is needed for this level of agreement.

To perform a comprehensive analysis taking account of 
the uncertainty of collected experimental data, 
a chi-squared analysis has been employed in this work, 
and the corresponding results of the $\chi^2/dof$ and \textit{p}-value 
are plotted in Fig. \ref{fig:stats}(b) and \ref{fig:stats}(c).
These figures give similar but more clear and 
credible results compared with Fig. \ref{fig:stats}(a).
The most meaningful result is that distinct improvements upon the $\chi^2/dof$ and 
\textit{p}-value are made by the QMCGGA form 
compared with the PHCGGA form.
Furthermore, the \textit{p}-values given by 
the calculations based on the FLAPW method and the QMCGGA approach
are 0.55, 0.93 and 0.92 for the subset C, D and E, respectively. 
The fact that these \textit{p}-values are larger than 0.05 
(related to 95\% CL) means that the corresponding deviations 
between the theoretical values and experimental data 
can be considered as the noise level of the dataset, 
so that a brand-new level of agreement is achieved for 
these materials with the theoretical lifetimes ranging from 
104 ps ($\tau_{\textbf{Ni}}$) to 287 ps ($\tau_{\textbf{CdTe}}$).
In other words,  
the calculations based on the FLAPW together with the new QMCGGA 
are adequately credible for predicting exact positron lifetimes for bulk materials 
even confronting them with repeated measurements.

Caused by the boosted confidence and that 
the best scheme (FLAPW \& QMCGGA) gives $\text{RMSD}_{min}=3.5$ ps, 
it is reasonable to conclude that the accuracy of these experimental values 
with a deviation from the best theoretical value 
being larger than 10 ps ($\approx3\times\text{RMSD}_{min}$) should be questioned. 
As a consequence, the experimental data of the following materials: 
Cr (120 ps), Sc (230 ps), Y (249 ps), TiO (140 ps), GaN (180 ps),
Pt (113.5 ps), Co\_alpha (118.5 ps), Tl (227.5 ps), TiC (157.5 ps) and MgO (150.75 ps)
should be strongly doubted and their corresponding calculated 
theoretical values are much more reliable.

But for materials with lower electron densities and larger lifetimes ($\gtrsim 300$ ps), 
the effectiveness of the best scheme is not proved. 
From left part of Fig. \ref{fig:ep-dt-FLAPW}, 
it is clear that the theoretical lifetimes for alkali-metal 
given by APGGA and PHCGGA approaches are 
in much better agreement with current experimental values. 
So, as shown in Fig. \ref{fig:stats}(a), 
the RMSD of QMCGGA related to the dataset A is not better than that of APGGA and PHCGGA. 
Meanwhile, taking into account the overestimation of the gradient correlation by APGGA 
prominently for semiconductors with larger $\epsilon^\star$, the less semiconductors  
a dataset involves, the better the APGGA behaves. 
This is consistent with our previous work 
where a limited dataset composed of 16 materials 
including alkali-metals and only three semiconductors is used \cite{Zhang14V2}. 
There, although the GGA form based on recent QMC simulation performs better for semiconductors, 
no remarkable improvement for the whole dataset is found.
When we take into consideration that the measured values for alkali-metals 
reported in 1967 are not suggested to be treated seriously \cite{Kurip14V89}, 
the benefit of the two QMC-GGA forms is swamped due to 
the mixture of reliable data and insufficient data. 
In addition, the positron density is able to exceed the electron density 
in the case of positron binding atoms and ions. 
Significantly, the almost exact annihilation rate 
calculated using many-body theory method for this atomic-type system 
\cite{Mitro02V65,Griba10V82,Green14V90} 
allows us to further validate these positron-electron correlation schemes. 
As mentioned above, 
the QMCLDA form gives a much larger enhancement factor in the high $r_s$ region 
compared with the reparameterized fQMCLDA form, 
although their discrepancy at low $r_s$ is minor. 
So, the distinct asymptotic behavior at large $r_s$ 
as well as the modified $\alpha$ 
would provide us an opportunity to establish a better DFT calculation scheme 
for this atomic-type system, and a revisit to this important question is valuable.

We further calculated the average $r_s$ sensed by positron 
defined by $r^{*}_{s} =\int r_s \cdot n^{e+} \cdot d\vec{r}$ 
based on the QMCGGA correlation scheme. 
The $r^*_s$ ranges from 1.6 (3.1) to 3.3 (4.9) 
for the materials in dataset A (alkali-metals). 
That means the assessment based on the results of dataset A is focused on 
the region $1.6 \lesssim r_s \lesssim 3.3$ and will be less affected by
the experimental data of alkali-metals. 
However, to determinate which approach is better suitable 
in the high $r^{*}_s$ (high lifetime) region
more measurements and exact many-body calculations in this field are needed.
Besides, it is easy to obtain that 
a change of 1 $\sigma_{exp}$ ps in any experimental value 
leads to a change of the order of 0.1 in the $\chi^2$ in general for dataset A, 
which presents the stability of our statistic results.

\section{Conclusions}
In summary, to probe the validity of several correlation schemes for 
the positron lifetime of bulk material, an original statistical analysis is implemented 
based on a more reliable experimental dataset with estimated measurement errors. 
Most significantly, the original GGA form (APGGA) is found overestimates 
the contribution of the gradient effect. 
Based on the FLAPW method for the most accurate electronic-structure calculations,  
the two GGA forms come from recent QMC calculations, 
especially the new GGA correlation schemes introduced in this paper, 
can significantly reduce the chi-squared into the 
95\% confidence region ({\it p}-value $>$ 0.05). 
This brand-new level of agreement demonstrates that the performance 
of theoretical prediction can be independence on the type of materials. 
Meanwhile, it is reasonable to state that several special experimental lifetimes 
should be questioned and the theoretical values are much more credible than ever. 
It should be noted that more accurate experiment data especially for 
those materials with larger lifetimes ($\gtrsim$ 300 ps) are needed 
for further progress on studies of material by using PAS.

\section*{Acknowledgments}
We would like to thank Rong-Dian Han and Wen-Zhen Xu for helpful discussions. 
And part of the numerical calculations in this paper have been done 
on the supercomputing system in the Supercomputing Center of 
University of Science and Technology of China.
This research was supported by National Natural Science Foundation of China 
(Grant Nos. 11175171 and 11105139).

\section*{Appendix A. All calculated theoretical lifetimes}
As listed in Table \ref{tab:all-theo-tau}, we present 
all calculated theoretical lifetimes for each bulk material by using 
nine different forms of LDA/GGA correlation schemes based on 
the accurate FLAPW electronic-structure calculations. 
For comparison, the corresponding average experimental 
values $\tau^{*}_{\text{exp}}$ are also listed.

\begin{table*} [htb]
\centering
\footnotesize
\caption{ \label{tab:all-theo-tau} 
All calculated theoretical lifetimes for each bulk material.
}
%
%
\begin{tabular}[b]{l p{13mm} p{13mm} p{13mm} p{13mm} p{14mm}  p{14mm} p{14mm} p{14mm} p{13mm} r}
\hline 
\hline
Material & BNLDA & APLDA  & APGGA  & PHCLDA & PHCGGA & fQMCLDA & fQMCGGA & QMCLDA & QMCGGA  & $\tau^{*}_{\text{exp}}$  \tabularnewline
\hline
\rule[.2ex]{0pt}{2ex} 
     Li & 299.75 & 259.63 & 283.10 & 277.43 & 295.00 & 304.67 & 316.77   & 304.55 & 318.82  & 291  \tabularnewline  
     Na & 328.18 & 290.71 & 337.51 & 309.87 & 341.26 & 339.00 & 359.06   & 345.45 & 370.63  & 338  \tabularnewline  
      K & 367.79 & 331.71 & 389.60 & 353.33 & 388.18 & 386.12 & 406.92   & 395.43 & 422.58  & 397  \tabularnewline  
     Rb & 383.52 & 349.95 & 410.41 & 372.35 & 407.69 & 406.25 & 426.97   & 416.03 & 443.06  & 406  \tabularnewline  
     Cs & 393.51 & 362.28 & 426.54 & 385.11 & 421.15 & 419.57 & 440.25   & 428.79 & 455.65  & 418  \tabularnewline  
     Ga & 187.53 & 166.60 & 200.66 & 176.31 & 193.36 & 190.81 & 200.18   & 187.15 & 198.35  & 198  \tabularnewline  
     Cr & 99.747 & 92.369 & 105.42 & 96.593 & 103.56 & 102.71 & 106.58   & 101.17 & 105.63  & 120  \tabularnewline  
     Sc & 194.82 & 171.04 & 191.48 & 181.49 & 193.11 & 197.19 & 203.96   & 191.92 & 199.93  & 230  \tabularnewline  
      Y & 214.27 & 186.85 & 207.47 & 198.68 & 209.95 & 216.57 & 223.02   & 211.02 & 218.80  & 249  \tabularnewline  
    TiO & 95.795 & 89.378 & 111.10 & 93.289 & 103.80 & 98.931 & 104.47   & 98.296 & 105.04  & 140  \tabularnewline  
    GaN & 144.75 & 125.42 & 161.99 & 131.73 & 149.00 & 140.97 & 150.05   & 139.37 & 150.54  & 180  \tabularnewline  
    ZnS & 216.22 & 178.81 & 240.94 & 189.17 & 218.07 & 204.61 & 219.82   & 202.13 & 221.01  & 230  \tabularnewline  
   ZnTe & 251.89 & 209.02 & 279.16 & 221.78 & 254.29 & 240.95 & 258.13   & 238.75 & 260.12  & 266  \tabularnewline  
   HgTe & 253.59 & 218.41 & 302.71 & 231.69 & 270.04 & 251.63 & 271.67   & 250.08 & 275.26  & 274  \tabularnewline  
   ZnSe & 231.15 & 190.72 & 257.34 & 201.99 & 232.90 & 218.85 & 235.14   & 216.53 & 236.75  & 240  \tabularnewline  
   PbSe & 207.92 & 181.12 & 232.67 & 191.91 & 216.03 & 208.07 & 220.83   & 204.67 & 220.50  & 220  \tabularnewline  
    NiO & 104.33 & 93.765 & 122.41 & 97.897 & 111.65 & 103.86 & 111.09   & 103.40 & 112.21  & 110  \tabularnewline  
     Be & 137.14 & 123.19 & 129.52 & 129.84 & 134.41 & 139.67 & 142.75   & 135.20 & 138.39  & 139.5  \tabularnewline  
     Pt & 95.970 & 89.265 & 104.70 & 93.244 & 100.82 & 98.994 & 103.02   & 97.865 & 102.67  & 113.5  \tabularnewline  
 Co\_a. & 96.700 & 89.925 & 106.58 & 93.920 & 102.69 & 99.692 & 104.53   & 98.665 & 104.25  & 118.5  \tabularnewline  
     Tl & 182.99 & 162.75 & 204.18 & 172.17 & 192.31 & 186.21 & 197.05   & 182.75 & 196.04  & 227.5  \tabularnewline  
    TiC & 107.86 & 99.384 & 115.16 & 104.08 & 111.97 & 110.92 & 115.18   & 109.03 & 114.13  & 157.5  \tabularnewline  
SiC\_3C & 140.57 & 122.30 & 144.06 & 128.62 & 139.06 & 137.91 & 143.47   & 135.05 & 141.82  & 139.0  \tabularnewline  
    GaP & 213.42 & 180.97 & 230.22 & 191.72 & 214.91 & 207.80 & 220.11   & 204.39 & 219.54  & 224.0  \tabularnewline  
   InAs & 240.84 & 205.59 & 270.54 & 218.12 & 248.27 & 236.93 & 252.87   & 234.58 & 254.42  & 252.0  \tabularnewline  
   InSb & 264.52 & 226.75 & 300.15 & 240.91 & 274.79 & 262.24 & 280.14   & 260.54 & 282.95  & 273.3  \tabularnewline  
   GaSb & 246.73 & 211.05 & 273.86 & 224.09 & 253.36 & 243.71 & 259.24   & 241.21 & 260.48  & 255.7  \tabularnewline  
     Bi & 224.36 & 197.64 & 245.48 & 209.88 & 232.53 & 228.29 & 240.40   & 224.82 & 239.87  & 235.5  \tabularnewline  
     Zr & 157.11 & 140.02 & 155.48 & 147.93 & 156.10 & 159.69 & 164.26   & 154.96 & 160.36  & 164.3  \tabularnewline  
  C\_d. & 95.935 & 86.675 & 103.07 & 90.456 & 98.178 & 95.909 & 99.933   & 95.103 & 100.10  & 106.5  \tabularnewline  
      W & 99.556 & 92.099 & 102.10 & 96.343 & 101.53 & 102.50 & 105.34   & 100.72 & 104.02  & 107.7  \tabularnewline  
     Pd & 104.08 & 96.328 & 116.49 & 100.75 & 110.76 & 107.16 & 112.51   & 105.87 & 112.27  & 110.5  \tabularnewline  
      V & 114.88 & 105.16 & 119.48 & 110.31 & 118.01 & 117.82 & 122.13   & 115.39 & 120.37  & 125.8  \tabularnewline  
SiC\_6H & 141.96 & 123.38 & 145.37 & 129.78 & 140.32 & 139.19 & 144.80   & 136.29 & 143.14  & 142.8  \tabularnewline  
    MgO & 127.67 & 108.58 & 145.91 & 113.67 & 131.90 & 121.06 & 130.78   & 120.30 & 132.23  & 150.8  \tabularnewline  
     Si & 213.98 & 182.01 & 217.38 & 193.12 & 210.01 & 209.83 & 218.87   & 205.12 & 216.17  & 218.1  \tabularnewline   
     Ge & 219.53 & 189.36 & 240.84 & 200.75 & 224.93 & 217.84 & 230.68   & 214.64 & 230.44  & 227.3  \tabularnewline  
     Mg & 230.40 & 200.07 & 217.33 & 213.02 & 224.59 & 232.64 & 240.11   & 227.60 & 236.08  & 228.6  \tabularnewline  
     Al & 164.67 & 145.78 & 154.17 & 154.25 & 159.60 & 166.90 & 170.26   & 161.31 & 164.97  & 163.9  \tabularnewline  
     Ti & 144.07 & 129.44 & 146.10 & 136.45 & 145.59 & 146.81 & 152.01   & 142.90 & 148.95  & 148.2  \tabularnewline  
     Fe & 100.86 & 93.423 & 109.26 & 97.679 & 106.08 & 103.84 & 108.50   & 102.48 & 107.85  & 109.8  \tabularnewline  
     Ni & 96.966 & 90.214 & 108.56 & 94.205 & 103.85 & 99.968 & 105.28   & 99.061 & 105.20  & 108.2  \tabularnewline  
     Zn & 137.25 & 124.32 & 148.82 & 130.76 & 143.87 & 140.22 & 147.59   & 137.56 & 146.07  & 150.4  \tabularnewline  
     Cu & 106.49 & 98.426 & 119.83 & 102.95 & 114.25 & 109.52 & 115.76   & 108.28 & 115.50  & 114.9  \tabularnewline  
     Nb & 121.29 & 110.42 & 122.98 & 115.99 & 122.46 & 124.17 & 127.72   & 121.13 & 125.31  & 121.6  \tabularnewline  
     Mo & 104.30 & 96.139 & 107.17 & 100.67 & 106.29 & 107.26 & 110.31   & 105.20 & 108.80  & 109.7  \tabularnewline  
     Ta & 115.90 & 105.92 & 117.89 & 111.15 & 117.47 & 118.80 & 122.30   & 116.11 & 120.16  & 120.0  \tabularnewline  
     Ag & 123.42 & 112.90 & 139.99 & 118.45 & 131.99 & 126.57 & 133.86   & 124.68 & 133.43  & 129.0  \tabularnewline  
     Au & 110.14 & 101.50 & 122.71 & 106.28 & 116.65 & 113.24 & 118.76   & 111.63 & 118.25  & 116.6  \tabularnewline  
     Cd & 157.17 & 141.16 & 173.14 & 148.87 & 165.10 & 160.27 & 169.16   & 157.22 & 167.89  & 176.8  \tabularnewline  
     In & 181.51 & 160.99 & 192.26 & 170.38 & 186.35 & 184.39 & 193.22   & 180.39 & 190.98  & 193.7  \tabularnewline  
     Pb & 188.73 & 167.05 & 201.90 & 176.93 & 194.10 & 191.71 & 201.02   & 187.62 & 198.96  & 201.4  \tabularnewline  
   GaAs & 222.20 & 189.47 & 244.71 & 200.81 & 226.68 & 217.81 & 231.53   & 214.85 & 231.76  & 228.9  \tabularnewline  
    InP & 235.23 & 198.85 & 259.03 & 210.91 & 238.96 & 229.01 & 243.84   & 226.18 & 244.63  & 242.8  \tabularnewline  
    ZnO & 155.31 & 131.63 & 184.37 & 138.20 & 162.79 & 147.81 & 160.64   & 147.08 & 163.04  & 160.4  \tabularnewline  
   CdTe & 274.20 & 226.73 & 312.64 & 240.69 & 279.99 & 261.68 & 282.31   & 260.33 & 286.24  & 286.8  \tabularnewline  
\hline 
\hline 
\end{tabular} 
\end{table*}

\end{document}